\documentclass[aps,apl,amsmath,amssymb,twocolumn]{revtex4}
\usepackage{graphicx}
\usepackage{graphics}
\usepackage{subfigure}
\usepackage{epstopdf}
\usepackage{color}
\usepackage{epsfig}
\usepackage{amssymb}
\usepackage{amsthm}
\usepackage{amsmath}
\usepackage{setspace}
\usepackage{longtable}
\usepackage{times}
\usepackage{epsfig,wrapfig}
\usepackage{dcolumn}   % needed for some tables
\usepackage{bm}        % for math
\usepackage{xcolor}

\begin{document}

\title{{Enhancement of perovskite solar cells by plasmonic nanoparticles}}

\author{Mikhail~Omelyanovich$^{1,2}$, Sergey~Makarov$^{2}$, Valentin~Milichko$^{2}$, Constantin~Simovski$^{1,2}$ }

\address{$^{1}$Department of Radio Science and Engineering, Aalto University, P.O. Box 13000, FI-00076, Aalto, Finland\\
	$^{2}$Laboratory of Metamaterials, University for Information Technology, Mechanics and Optics (ITMO), St. Petersburg 197101, Russia\\}

\begin{abstract}
Synthetic perovskites with photovoltaic properties open a new era in solar photovoltaics. Due to high optical absorption perovskite-based thin-film solar cells are usually considered as fully absorbing solar radiation on condition of ideal blooming. However, is it really so? The analysis of the literature data has shown that the absorbance of all photovoltaic pervoskites has the spectral hole at infrared frequencies where the solar radiation spectrum has a small local peak. This absorption dip results in the decrease of the optical efficiency of thin-film pervoskite solar cells by nearly 3\% and close the ways of utilise them at this range for any other applications. In our work we show that to cure this shortage is possible complementing the basic structure by an inexpensive plasmonic array.


\end{abstract}

\maketitle

\section{Introduction}

Synthetic perovskites are materials with chemical formula $ABX_3$ whose molecular crystal lattices are cubic, orthorhombic and tetragonal \cite{Perov}.
Element A is an organic molecule called large cation ($CH_3NH_3^+$, $NH_2CH=NH_2^+$, etc.). Element B, called small cation, is a heavy metal (usually, Pb, Ge, Eu, Cu, or Sn).
Element X can be I, Br, Cl or F (thus, $BX_3$ is a metal halide). Perovskites are polycrystal structures with several lattice structures which can be within one domain coordinated in different crystal planes. Therefore, perovskite compounds are mechanically amorphous \cite{Perov}. Their synthesis is a cheap low-temperature chemical process \cite{Perov}. Initially, perovskites were used in solar photovoltaics as nanoparticle counterparts of the dye molecules in dye-sensitized solar cells \cite{1}. Since that initial work, where the record (for that time) efficiency was obtained for a dye-sensitized solar cell, the efficiency of perovskite-based solar cells has grown by an order of magnitude \cite{2}. This huge progress achieved in 6 years allows researchers to claim a new era in solar photovoltaics when the solar energy is becoming a really mass product due to perovskites \cite{3}.
Perovskites are exitonic photovoltaic materials like organic dyes or quantum dots. However, unlike them they also possess the rather noticeable conductance for both electrons and holes \cite{2,3,4}. Moreover, due to a broad spectrum of direct p-p optical transitions they possess very high optical absorption \cite{2,3,4,Review}. Therefore, in a truly efficient  perovskite-based solar cell the perovskite layer is sandwiched between the hole-conducting and electron-conducting materials matched to the current-collecting electrodes \cite{2,3,Review}. The optimal thickness of this photovoltaic layer depending on the explicit design is within 400-500 nm. In other words, the modern perovskite-based solar cell is a thin-film solar cell \cite{Review}. This peculiarity shares out the perovskite solar cells from the raw of thin-film solar cells, where the micron or submicron thickness of the photovoltaic layer is dictated by the economy, compatibility with the roll-to-roll technology, flexibility, etc. Perovskite solar cells should be thin in order to be efficient. To make the perovskite layer thicker than 500 nm is not reasonable due to the strong increase of the recombination and, especially, ohmic losses across the layer. Smaller thickness than 400 nm would result in the abrupt increase of the parasitic capacitance which worsens the fill-factor of the solar cell. Moreover, such the decrease of the thickness results in multiple parasitic connections between the hole-conducting and electron-conducting materials through the pores in the perovskite layer (as an any other mechanically amorphous material perovskite is porous on the nanoscale). These connections result from the diffusion of amorphous materials. They shunt the photoinduced current and suppress the open-circuit voltage of the solar cell.

\begin{figure}[ht]
	\centering
	\includegraphics[width=8cm]{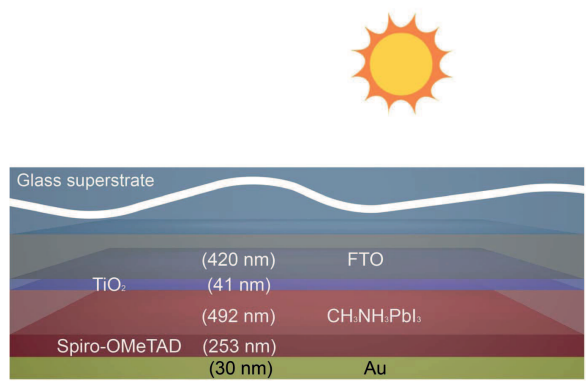}
	\caption{Schematics of the practical heterojunction configuration of a high-efficiency perovskite-based solar cell in accordance to \cite{Base}. The anti-reflecting coating on top of the superstrate is not shown.} \label{fig1}
\end{figure}

Thus, the optimal efficiency of the photovoltaic layer in modern perovskite solar cells is selected within 400-500 nm not from  the requirement of the optical efficiency.
However, the strong optical absorption makes this thickness also favorable for highest optical efficiency. The solar light transmitted to the perovskite layer through the anti-reflecting coating
is mainly absorbed before it attains the bottom electrode. It is enough to suppress the reflection in order to achieve the maximal optical efficiency. This maximal efficiency on condition of the ideal blooming is restricted only by losses in the top (transparent) electrode \cite{Review,4}. The integral (over the solar spectrum) power loss in the top electrode of the multilayer structure can be easily calculated analytically. For the explicit structure shown in our Fig.~1 (see \cite{Base}), the top electrode is a submicron layer of FTO (fluorine-doped tin oxide), and the integral power loss in it is nearly equal to 5\%. Then, in accordance to \cite{4}, the photovoltaic absorption efficiency of the solar cell should be equal to 95\%. An imperfectness of the anti-reflecting results in some unavoidable reflection losses which can be estimated for the best realistic coatings as 10\%. Then we obtain the total optical efficiency of the solar cell 85\% (10\% for the integral reflection power loss  is a realistic estimate for the so-called moth-eye full-angle anti-reflecting coating \cite{5}. In principle, it is possible to reduce the reflection losses to 1-2\% using multilayer submicron coatings, however, they are very expensive and not suitable for solar cells dedicated for the domestic power supplies \cite{7}).

This estimation of the optical efficiency of an optimized modern perovskite solar cell within the interval 80-90\% (it can be found also in \cite{6}) would be a very good result (if correct). Other thin-film solar cells with the similar thickness of the photovoltaic layer (e.g. those based on the amorphous silicon or epitaxial silicon) may attain the so high values of the optical efficiency only when they are furnished by so-called light-trapping structures (LTSs). This is so because both amorphous silicon and epitaxial silicon layers of typical thickness 300-500 nm do not sufficiently absorb the incident solar light. A significant part of solar radiation attains the bottom electrode and results in so-called transmission losses. Transmission losses are especially important for flexible (amorphous) solar cells in which the bottom electrode cannot be a light reflector. To suppress these transmission losses one needs to trap the light inside the photovoltaic layer (see e.g. \cite{7,8,9}). Light trapping is the linear operation  which prevents the propagation of the light across the photovoltaic layer. Instead, the incident wave power is spent either to the excitation of the waveguide modes in the photovoltaic layer (then the LTS operates like an optical facet) or to the creation of plasmonic hot spots in this layer (then the LTS operates like a nanofocusing device).

In absence of the LTSs usual thin-film solar cells have the optical efficiency within the interval 40-50\%. In presence of the best known LTS, performed as regular arrays of multi-resonant nanoantennas, the optical efficiency may attain 80-90\% \cite{10}, however, the cost of such solar cells will increase dramatically.  There is a body of literature (see e.g. \cite{11,12,13}) where the authors claim the suppression of the transmission losses by rather inexpensive LTSs, performed as random plasmonic (silver or gold) arrays chemically grown (usually) beneath of the photovoltaic layer or even on top of it. However, to our knowledge, no one of these random arrays stands the fair comparison with the anti-reflecting coating. This is so because plasmonic metals -- silver and gold -- are quite lossy in the visible range of wavelengths. One half of the plasmonic hot spot is located in the host materials, however, another half of the hot spot is inside the metal. Therefore, the parasitic absorption in such the LTS turns out to be comparable with that in the photovoltaic material. In other words, for such solar cells the parasitic transmission losses in the bottom electrode are substituted by similar parasitic losses in the LTSs, that evidently makes the light trapping meaningless \cite{14}. The situation with plasmonic LTSs for thin-film solar cells is as follows: regular arrays can be efficient but too expensive, random arrays can be inexpensive but not efficient.

A great advantage of the perovskite compared to silicon is its high optical absorption which seemingly allows the thin-film solar cell to avoid transmission losses. As we have mentioned, it is practically believed that the absorption efficiency of the perovskite solar cell differs from 100\% only due to optical losses in the top electrode (see e.g. in \cite{Review,4,6}). However, it is, not really correct. In accordance to the literature data, see e.g. in \cite{Base}, photovoltaic perovskites possess high optical absorption not at all wavelengths of the solar spectrum.  The issue of transmission losses is not fully avoided for these solar cells. The absorption coefficient of photovoltaic perovskites has two spectral holes -- at 800-920 nm and at 950-1050 nm, where the infrared part of the standard solar spectrum on the Sea level has two small but noticeable local maxima. This shortage decreases the absorption efficiency, at least slightly. For the moment, this slight decrease is mostly neglected by researchers dealing with perovskite solar cells because they are occupied by more actual problems, such as the chemical stability of involved materials, etc. (see, e.g. in \cite{Review}), but some of them offer tandem structures\cite{ReviewHybrid} or submicron patterning\cite{submicronPatterning} as a solution of the spectral hole problem. In the present paper, we show how to do it using plasmonic nanoparticles. In addition with available laser equipment it will be much cheaper than both previously proposed solutions.

\section{Problem formulation and suggested solution}
\label{sec:Basic}

Consider the structure presented in Fig.~\ref{fig1}, assuming that on top of the superstrate performed as an organic glass layer there is an ideal anti-reflecting coating. Then the incident wave in the whole solar spectrum is fully transmitted into the glass. This assumption is needed to simplify our full-wave numerical simulations. In this simulations we assume that the wave is incident from the half-space of glass to the top interface of the structure (glass-FTO). For our purposes, it is enough to consider the normal incidence of the solar light. The dimensions of the simulated structure are depicted in Fig.~\ref{fig1} and correspond to the practical solar cell \cite{Base}. Namely, the FTO layer is 420 nm thick, 41 nm is the thickness of the titan dioxide layer, 492 nm is the thickness of the perovskite ($CH_3NH_3^+PbI_3$) layer, and the hole-conducting layer of Spiro-OMeTAD is 253 nm thick. The golden back contact has the thickness 30 nm. The properties of the mechanic carrier of the whole structure do not play any role, since the incident wave does not practically transmit through this structure.

\begin{figure}[ht]
\centering
\includegraphics[width=\linewidth]{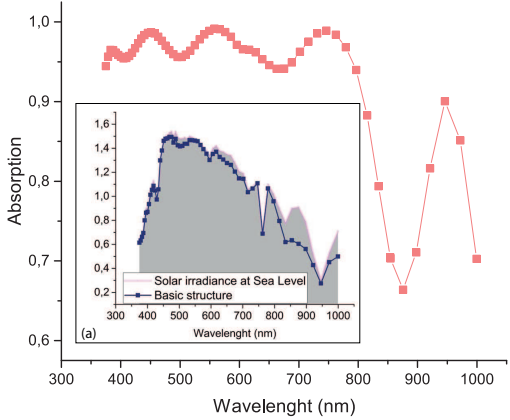}
\caption{Absorption coefficient of the perovskite layer in the structure shown in Fig. 1 versus wavelength. In the inset (a) we show the solar spectrum AM 1.0 in arbitrary units (shadowed) and the spectrum of absorption in the perovskite layer.} \label{fig2}
\end{figure}

The absorption in the layers of the structure has been analyzed using the CST Studio software confirmed by the analytical calculations of the multilayer structure via transmission matrices of layers (see e.g. in \cite{Mikhail}). Material parameters of all these layers have been taken from \cite{Base}. In Fig. \ref{fig2} we show the simulated coefficient of the photovoltaic absorption (percentage of the incident power spectrum absorbed in the perovskite layer) over the whole solar spectrum. We see two significant dips of absorption centered at 870 and 1000 nm. In the inset Fig.~\ref{fig2}(a) we show the spectrum of radiation absorbed in the perovskite -- solar irradiance spectrum (shown as a in shadow this inset) multiplied by the photovoltaic absorption coefficient.
This spectrum implies that 3\% of the solar radiation transmitted to the perovskite is lost.

This power transmits through the pervoskite layer and is partially absorbed in the hole-conducting layer. A significant part of this power is reflected back to the glass superstrate. Nearly 2\% of the solar power is lost in the range 820-900 nm  where the local maximum occurs at 870 nm. In this work we will show how to avoid this 2\% loss.

As it was already mentioned, the subwavelength concentration of the local electric field within a plasmonic hot spot enhances the absorption around the plasmonic nanoparticle. On the other hand, we have noticed that the harmful absorption inside the nanoparticle is also enhanced. However, this second harmful effect is critical for golden nanoparticles operating in the visible range. In the visible range the figure of merit (FOM) of the optical constant $\varepsilon_G$ of gold is not sufficiently high due to the interband optical transitions. In the infrared range this FOM is much higher. If we manage to locate the plasmon resonance of a golden nanoparticle on the wavelengths 800-900 nm, the distribution of the power lost in the internal and external parts of the plasmonic hot spot will be very different from the situation analyzed in \cite{14}. No interband transitions exist in this range where the Drude model for $\varepsilon_G$ is practically exact \cite{gold}. The skin-depth of gold in this band is nearly 20 nm, and if the particle is substantially larger (say, 100-200 nm thick) the electric field weakly penetrates through its surface. The hot spot will be located mainly in the host medium. Though the absorption coefficient of the perovskite is small in this range, the losses in the host perovskite layer may dominate over the losses in gold.  The location of the plasmon resonance in the band 700-1000 nm is possible due to the high refractive index $n_P$ of the perovskite. It is easy to check for a golden nanosphere where the condition of the individual plasmon resonance of a nanosphere in the host with permittivity $\varepsilon_P$ is, in accordance to the Drude model, as follows:
$$
{\rm Re} (\varepsilon_G)\approx -{\lambda^2\over \lambda_{pG}^2}=-3{\rm Re} (\varepsilon_P)\approx -3n_P^2.
$$
Here $\lambda_{pG}=$138 nm is the plasma wavelength for gold \cite{gold}. This equation gives for the wavelengh $\lambda$ of the individual plasmon resonance the approximate result $\lambda\approx $750 nm. The exact location of the resonance on the frequency axis is mediated by the density of the array of golden spheres \cite{gold111}.

To check this idea we need to submerge substantial golden nanospheres into perovskite. If these submerged nanospheres are located in the top part of the perovskite layer, their harmful action will be probably dominating. Really, beyond the useful plasmon resonance these metal particles will reflect the light before it is absorbed by the perovskite. Moreover, in the visible range these substantial particles possess high-order (multipole) plasmon resonances at which the light will be absorbed in the metal because in the visible range gold is an essentially lossy material. Therefore, we have to submerge our particles to the bottom part of the perovskite layer. Then we will avoid parasitic effects and keep the useful one. Within the spectral hole of the perovskite absorption the light will transmit through the layer, attain our plasmonic array, form the hot spots around our nanoparticles and absorb in these spots.

Of course, our array should be prepared in a cheap enough way: the maximally possible 2\% increase of the optical efficiency may justify only similarly low increase of the fabrication costs.
Below we will discuss these technological issues. In the present section, we only describe the optimal parameters of the plasmonic structure. Besides of solid golden nanospheres, we also
made simulations for golden nanoshells -- such nanoparticles are available on the market. The basic structure is the same as shown in Fig.~\ref{fig1}. Nanoparticles are inserted inside the perovskite layer to a certain optimal depth $f$. This depth was found after a set of numerical simulations in CST Studio. In our simulations the array was assumed to be periodic in order to save the computational time: the periodicity allows us to reduce the array electromagnetic problem to the cell problem. The periodicity is not significant for our effect. What is significant is the proper mean inter-particle distance. Notice, that if this distance is large the near-field interaction of nanoparticles vanishes and the collective plasmon resonance degenerates into the individual plasmon resonance which happens at $\lambda=$750 nm. Simulations have shown: in order to place the resonance to $\lambda=$850 nm the mean distance between the centers of adjacent particles should be equal 430 nm.

\section{Numerical results and discussion}
\label{PV}

\begin{figure}[ht]
	\centering
\subfigure[]{\includegraphics[width=0.55\linewidth]{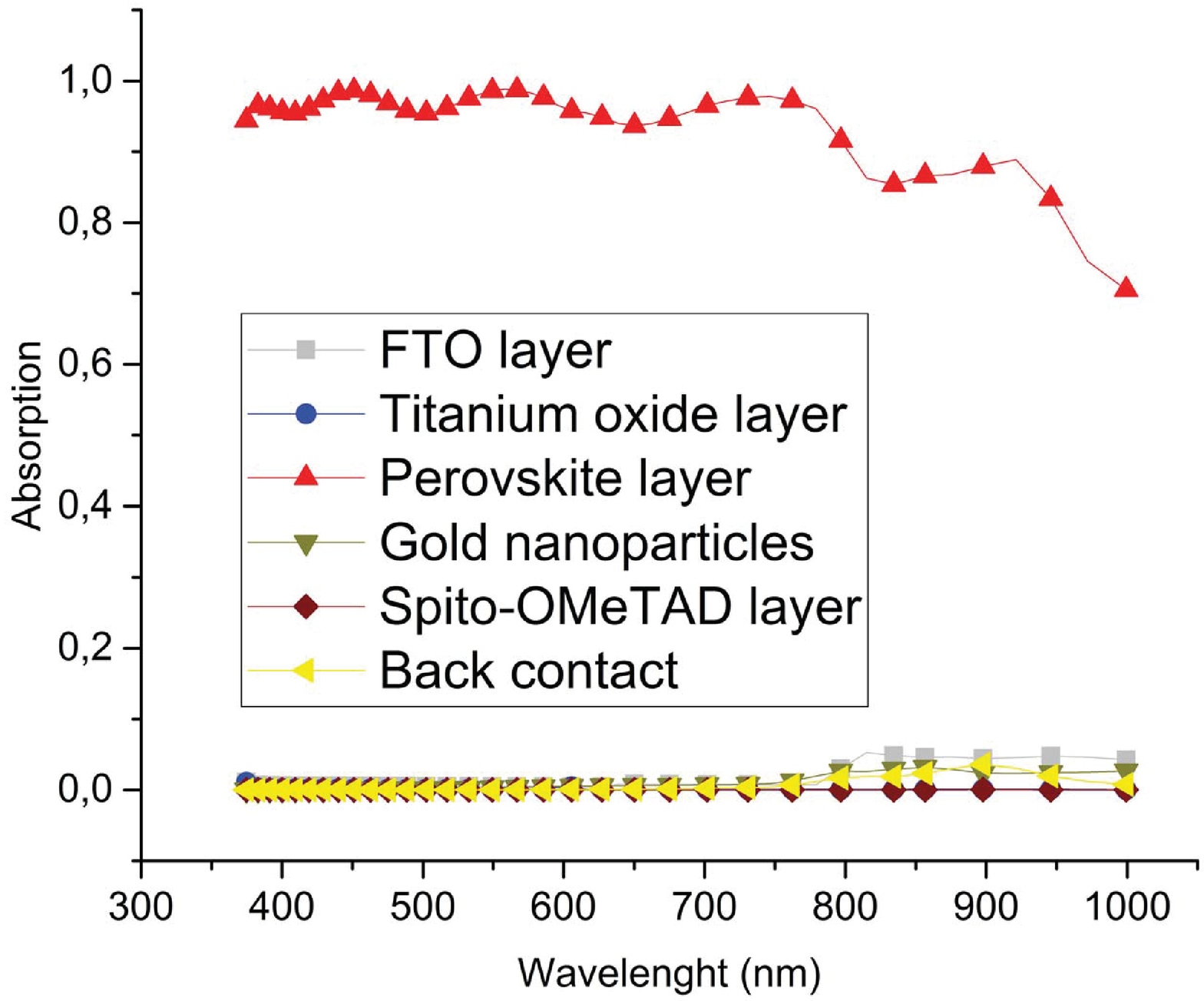}}
\subfigure[]{\includegraphics[width=0.4\linewidth]{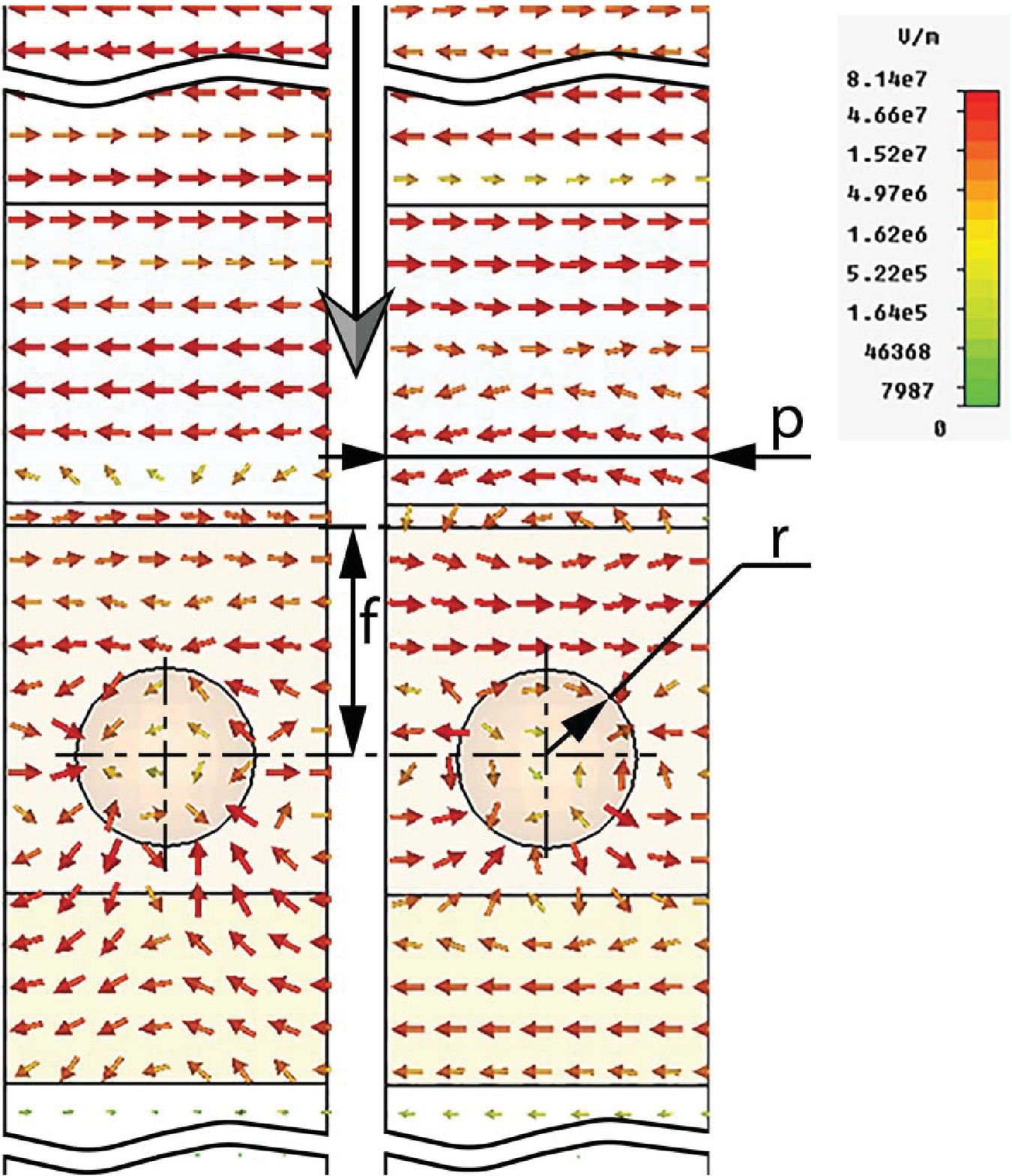}}
	\caption{Absorption of all components of our structure (a). (b) -- Vector color maps of the electric field at $\lambda=$850 nm: zero phase (left panel) and 90$^\circ$ phase (right panel). } \label{fig3}
\end{figure}

In Fig.~\ref{fig3} we depict the partial absorption coefficients corresponding to all components of our structure with $p$ -- period of the structure -- 430 nm. All these parameters are taken from  \cite{Base}. As to golden nanospheres, their radius was chosen $r$=120 nm, and $f$ -- distance from the center of the nanospheres to the boundary of the TiO$_2$ layer -- equals to 309 nm.

\begin{figure}
	\centering
	\includegraphics[width=\linewidth]{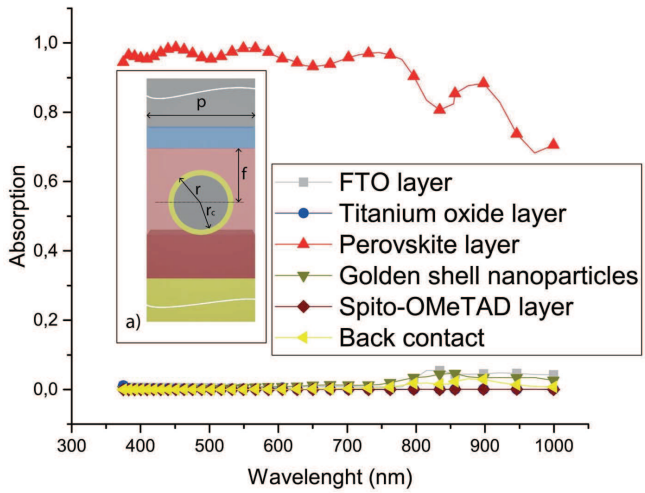}
	\caption{Absorption of the components of the structure shown in the inset (a).} \label{fig4}
\end{figure}

Compared to the results depicted in Fig.~\ref{fig2} the absorption in the perovskite layer at 850-920 nm increases due to the presence of nanoparticles from 0.65-0.75 to 0.82-0.85, whereas the absorption of all other components of the structure, including nanoparticles, keeps negligibly small. In Fig.~\ref{fig3}(b) we show two vector color maps of the local electric field corresponding to two time moments shifted from one another by one quarter of the period. These color maps are typical for collective plasmon resonances (see similar maps in \cite{gold1}). This fact confirms our guess that the periodicity of the array is not significant. Another observation which allows us to conclude that the observed resonance is not sensitive to the periodicity is the dynamics of the resonance wavelength versus the period $p$, as we have already mentioned. Like in usual plasmonic arrays utilized to enhance the photovoltaic absorption in thin-film solar cells \cite{gold111} the strict periodicity is not needed in our case.

\begin{figure}
	\centering
	\includegraphics[width=\linewidth]{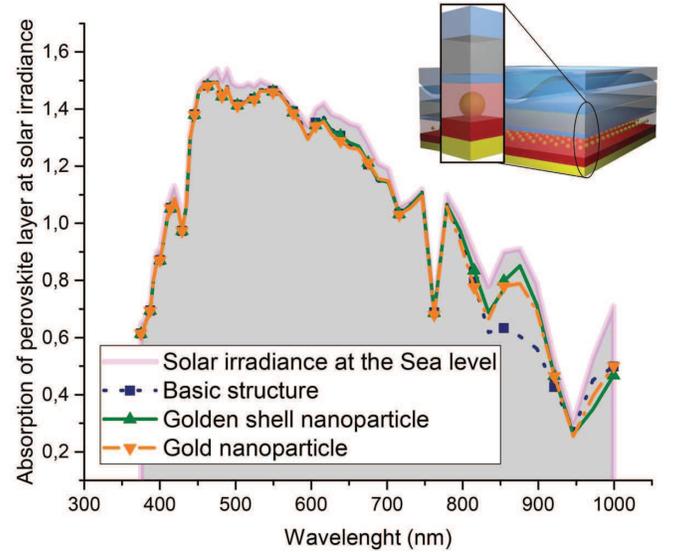}
	\caption{Absorption in the perovskite layer for several variants of our solar cell (basic structure, the structure enhanced by solid nanospheres, and that enhanced by nanoshells) taking into account the solar irradiance spectrum.} \label{fig5}
\end{figure}

For golden nanoshells we have done the similar set of simulations keeping the same overall radius $r=$120 nm of the particle and assuming the silica core of radius $r_c$=90 nm inside the shell. For  this case we have obtained better results. Definitely, the plasmon resonance frequency keeps the same for the same period of the array. However, decreasing the amount of gold we increase the percentage of the power absorbed in the perovskite layer. The results of this study are illustrated by Fig.~\ref{fig4}.

\begin{figure}
	\centering
	\includegraphics[width=0.8\linewidth]{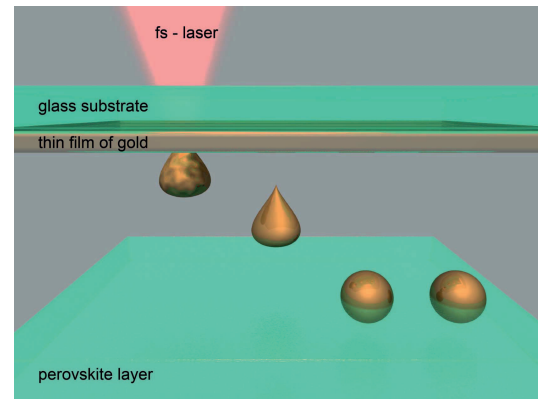}
	\caption{A sketch of the laser-induced forward transfer of golden nanoparticles into perovskite.} \label{fig6}
\end{figure}

Fig.~\ref{fig5} shows the absorption spectrum in the perovskite layer in comparison with the solar irradiance spectrum in the range 300-1000 nm.
The area under the absorption curve normalized to the area under the curve of the solar irradiance is the absorption efficiency.
This value for the basic structure is nearly equal 92\% (5\% is lost in FTO). For the structure enhanced by solid nanospheres it is equal 93.2\%, and for nanoshells it is 93.8\%.
The absorption efficiency 95\% cannot be approached even using nanoshells because the second spectral hole in the perovskite absorption holds at $\lambda=$1000 nm.
To address this problem we may broaden the plasmon resonance of the array so that it would cover both regions of weak absorption: 800-920 nm and 980-1020 nm.
However, in the present work we only aim to show that the plasmon resonances of golden nanoparticles may really enhance the perovskite solar cell.
All we need to finalize the present study is to explain how to prepare our structure in a cheap way.

We have already mentioned that to justify the 2\% gain in the optical efficiency the fabrication cost of our plasmonic array has to be low. We hope that it will be low if we use the
femtosecond-laser-induced forward transfer technique -- LIFT technique \cite{gold2,gold3}. LIFT is a high-speed and low-cost technology which simultaneously produces metal or even dielectric \cite{gold4,gold5} nanoparticles and transfers them to the target host. The host should be either soft like polymers or porous like perovskites. The procedure is illustrated by Fig.~\ref{fig6}. The laser locally melting the golden film is assumed in this picture moving from the right to the left. Both averaged inter-particle distance and particle size can be independently controlled by laser focusing conditions and the energy in the laser pulse. The depth of the embedding into the target substrate depends on the distance to it from the Au film. It is worth noticing that both the temperature and velocity of the embedded nanoparticle are also controllable \cite{gold4}, making possible the transfer of either hot or cold nanoparticles.
The perovskite solar cell shown in Fig.~\ref{fig1} is produced from top to bottom, i.e. prepared on the glass substrate (which can be later reduced to the optimal thickness and furnished by the anti-reflecting coating). Therefore, before covering the perovskite with the Spiro-OMeTAD one may put on the perovskite a glass plate covered by gold and to apply the LIFT technique.
After submerging the nanoparticles into perovskite, the glass plate can be easily removed, and the fabrication continues.
Of course, this method is not directly suitable for embedding the nanoshells. However,
we believe that the laser ablation of multilayer films could be realized for nanoshells as well.
To our opinion, LIFT is a promising candidate to fabricate plasmon-enhanced solar cells of 3d generation -- besides of pervoskite solar cells there are several
solar cells in which irregular arrays of plasmonic particles may enhance the overall efficiency due to different physical mechanisms \cite{Plasmon1,Plasmon2, Plasmon3}.

\section{Conclusions}

In the present paper we have proposed and studied a plasmonic array which allows a thin-film perovskite solar cell to fully absorb the whole solar spectrum from ultraviolet up to wavelength 1000 nm. The problem of transmission losses in the band 820-950 nm is theoretically solved embedding golden nanoparticles into the bottom part of the pervoskite layer. Nanoparticles located there increase the useful absorption in the infrared spectral hole of the perovskite by 32\% and do not distort the photovoltaic absorption at other wavelengths. This soluthion also makes possible for perovskite solar cells to operate as NIR photodetectors. As to basic(solar cell) application, the gain 1.8\% may be sufficient to justify the fabrication costs because the suggested (LIFT) technology is low-cost and high-speed, i.e. compatible with the large-area fabrication principle.


\begin{thebibliography}{99}

\bibitem{Perov}
A. Ecija, K. Vidal, A. Larranaga, L. Ortega-San-Martín, and M.I. Arriortua,
Synthetic Methods for Perovskite Materials –- Structure and Morphology, in: Advances in Crystallization Processes, Y. Mastai, Ed., InTech Publisher, Rieka--Shanghai, 486--506 (2012)

\bibitem{1}
A. Kojima, K. Teshima, Y. Shirai, T. Miyasaka, Organo-metal halide perovskites as visible-light sensitizers for photovoltaic cells. J. Am. Chem. Soc.
\textbf{131}, 6050–-6051 (2009)

\bibitem{2}
H.S. Jung and N.-G. Park, Perovskite solar cells: From materials to devices, Small \textbf{11}, 10--25 (2015)

\bibitem{3}
B. Wang, X. Xiao, T. Chen, Perovskite photovoltaics: a high-efficiency newcomer to the solar cell family, Nanoscale \textbf{6}, 12287--12297 (2014)

\bibitem{4}
Q, Shen, Y. Ogomi, J. Chang, T. Toyoda, K. Fujiwara, K. Yoshino, K. Sato, K. Yamazaki, M. Akimoto, Y. Kuga, K. Katayamad, and S. Hayase, Optical absorption,
charge separation and recombination dynamics in Sn/Pb cocktail perovskite solar cells and their relationships to photovoltaic performances,
European Journal of Chemical Physics and Physical Chemistry \textbf{15}, 1062--1069 (2014)

\bibitem{Review}
M.-E. Ragoussia and T. Torres, New generation solar cells: concepts, trends,
Chem. Commun. \textbf{51}, 3957--3972 (2015)

\bibitem{ReviewHybrid}
Qi Chena, Nicholas De Marcoa, Yang (Michael) Yanga,
Tze-Bin Songa, Chun-Chao Chena, Hongxiang Zhaoa, Ziruo Honga, Huanping Zhoua, Yang Yanga, The organic-inorganic hybrid halide perovskite for optoelectronic applications, Nano Today \textbf{10}, 355--396 (2015)

\bibitem{submicronPatterning}
Mohd S. Alias, Yang Yang, Tien K. Ng, Ibrahim Dursun, Dong Shi, Makhsud I. Saidaminov, Davide Priante, Osman M. Bakr, and Boon S. Ooi, Enhanced Etching, Surface Damage Recovery, and Submicron Patterning of Hybrid Perovskites using a Chemically Gas-Assisted Focused-Ion Beam for Subwavelength Grating Photonic Applications, J. Phys. Chem. Lett., 7, 137--142 (2016)
{DOI: 10.1021/acs.jpclett.5b02558}


\bibitem{5}
J. Tommila, A. Aho, A. Tukiainen,
Moth-eye antireflection coating fabricated by nanoimprint lithography on 1 eV dilute nitride solar cell,
Prog. Photov. Res. Appl. \textbf{21} 1158--1162 (2012)

\bibitem{6}
H.J. Snaith, Perovskites: the emergence of a new era for low-cost, high-efficiency solar cells, J. Phys. Chem. Lett. \textbf{4}, 3623--3630 (2013)

\bibitem{7}
S.B. Mallick, M. Agrawal, and P. Peumans, Optimal light trapping in ultra-thin photonic crystal crystalline silicon solar cells, Opt. Express 18: 5691 (2007)

\bibitem{8}
N.C. Panoiu and R.M. Osgood, Enhanced optical absorption for photovoltaics via excitation of waveguide and plasmon-polariton modes, Opt. Lett. 32: 2825-2829 (2007)

\bibitem{9}
H.A. Atwater and A. Polman, Plasmonics for improved photovoltaic devices, Nature Mat. 9: 205-209 (2010)

\bibitem{10}
P.A. Spinelli, V.E. Ferry, and J. van de Groep, Plasmonic light trapping in thin-film Si solar cells, J. Optics \textbf{14}, 024002 (2012).

\bibitem{11}
K.R. Catchpole and A. Polman, Design principles for particle plasmon enhanced solar cells, Appl. Phys. Lett. \textbf{93}, 191113 (2008)

\bibitem{12}
T.V. Pfeiffer, J. Ortiz-Gonzalez, R. Santbergen, H. Tan, A. Schmidt Ott, M. Zeman, and A.H.M. Smets, Plasmonic nanoparticle films for solar cell applications fabricated by size-selective aerosol deposition, Energy Procedia \textbf{60}, 3-–12 (2014)

\bibitem{13}
H. Tan, R. Santbergen, A.H.M. Smets, and M. Zeman,
Plasmonic light trapping in thin-film silicon solar cells with improved self-assembled silver nanoparticles, Nano Lett. \textbf{12}, 4070−-4076 (2012)

\bibitem{14}
Yu. A. Akimov, W. S. Koh, S.Y. Sian, and S. Ren, Nanoparticle-enhanced thin-film solar cells: metallic or dielectric nanoparticles? Appl. Phys. Lett. \textbf{96}, 073111 (2010)

\bibitem{Base}
J.M. Ball, S.D. Stranks, M.T. H{\"o}rantner, S. H{\"u}ttner, W. Zhang, E.J.W. Crossland, I. Ramirez, M. Riede, M.B. Johnston, R.H. Friend and H.J. Snaith,
Optical properties and limiting photocurrent of thin-film perovskite solar cells, Energy Environ. Sci., \textbf{8}, 602 (2015)

\bibitem{Mikhail}
M. Omelyanovich, Y. Ra'di and C. Simovski,
Perfect plasmonic absorbers for photovoltaic applications,
Journal of Optics, \textbf{17}, 125901 (2015)

\bibitem{gold}
A. Pinchuk, U. Kreibing, A. Hilger,
Optical properties of metallic nanoparticles: inﬂuence of interface effects and interband transitions,
Surface Sci. \textbf{557}, 269--280 (2004)


\bibitem{gold111}
S. Pillai, K.R. Catchpole, T. Trupke, and M.A. Green, Surface-plasmon enhanced solar cells,
J. Appl. Phys. \textbf{101}, 093105 (2007)

\bibitem{gold1}
M.V. Bashevoy, V.A. Fedotov and N.I. Zheludev, Optical whirlpool on an absorbing
metallic nanoparticle, Opt. Express \textbf{13}, 8372--8379 (2005)

\bibitem{gold11}
Z.B. Wang, B.S. Luk’yanchuk, M.H. Hong, Y. Lin, and T.C. Chong,
Energy ﬂow around a small particle investigated by classical Mie theory,
Phys. Rev. B \textbf{70}, 035418 (2004)



\bibitem{gold2}
D. A. Willis and V. Grosu,
Microdroplet deposition by laser-induced forward transfer,
Appl. Phys. Lett. \textbf{86}, 244103 (2005).

\bibitem{gold3}
A.I. Kuznetsov, A.B. Evlyukhin, M.R. Goncalves, C. Reinhardt, A. Koroleva, M.L. Arnedillo, R. Kiyan, O. Marti, and B.N. Chichkov,
Laser fabrication of large-scale nanoparticle arrays for sensing applications,
ACS Nano \textbf{5}, 4843 (2011)

\bibitem{gold4}
U. Zywietz, A. B. Evlyukhin, C. Reinhardt, and B. N. Chichkov,
Laser printing of silicon nanoparticles with resonant optical electric and magnetic responses,
Nature Comm. \textbf{5}, 3402 (2014)

\bibitem{gold5}
P. Dmitriev, S.V. Makarov, V. Milichko, I. Mukhin, A. Gudovskikh, A. Sitnikova, A. Samusev, A. Krasnok, and P. Belov,
Laser fabrication of crystalline silicon nanoresonators from an amorphous film for low-loss all-dielectric
nanophotonics, Nanoscale (2015), DOI: 10.1039/C5NR06742A

\bibitem{Plasmon1}
C. Clavero, Plasmon-induced hot-electron generation at nanoparticle/metaloxide interfaces for photovoltaic and photocatalytic devices,
Nat. Photonics \textbf{8}, 95–-103 (2014)

\bibitem{Plasmon2}
H. Chalabi and M. L. Brongersma, Plasmonics: harvest season for hot electrons, Nat. Nanotechnol. \textbf{8}, 229–-231 (2013)

\bibitem{Plasmon3}
F. Pastorelli, S. Bidault, J. Martorell, N. Bonod, Self-assembled plasmonic oligomers for organic photovoltaics, Adv. Opt. Mat. \textbf{2}, 171--175 (2014)

\end{thebibliography}
\end{document}